# The Optical and Mechanical Design for the 21,000 Actuator ExAO System for the Giant Magellan Telescope: GMagAO-X

Laird M. Close[a], Jared R. Males[a], Olivier Durney[a], Fernando Coronado[a], Sebastiaan Y. Haffert[a], Victor Gasho[a], Alexander Hedglen[b,a], Maggie Y. Kautz[b,a], Tom E. Connors[a], Mark Sullivan[a], Olivier Guyon[a,b,c,d], Jamison Noenickx[a]

[a]Steward Observatory, University of Arizona, USA
[b]James C. Wyant College of Optical Sciences, University of Arizona, USA
[c]Subaru Telescope, National Astronomical Observatory of Japan
[d]Astrobiology Center, National Institutes of Natural Sciences, Japan

## ABSTRACT

GMagAO-X is the ExAO coronagraphic instrument for the 25.4m GMT. It is designed for a slot on the folded port of the GMT. To meet the strict ExAO fitting and servo error requirement (<90nm rms WFE), GMagAO-X must have 21,000 actuator DM capable of ≥2KHz correction speeds. To minimize wavefront/segment piston error GMagAO-X has an interferometric beam combiner on a vibration isolated table, as part of this "21,000 actuator parallel DM". Piston errors are sensed by a Holographic Dispersed Fringe Sensor (HDFS). In addition to a coronagraph, it has a post-coronagraphic Low Order WFS (LLOWFS) to sense non-common path (NCP) errors. The LLOWFS drives a non-common path DM (NCP DM) to correct those NCP errors. GMagAO-X obtains high-contrast science and wavefront sensing in the visible and/or the NIR. Here we present our successful externally reviewed (Sept. 2021) CoDR optical-mechanical design that satisfies GMagAO-X's top-level science requirements and is compliant with the GMT instrument requirements and only requires COTS parts.

## 1. INTRODUCTION

A key reason we are pursuing the difficult effort of construction of the ELT-class of telescopes is for the characterization of habitable planets. Indeed, the recent *Astro2020* decadal survey listed this as the number one science goal for the "Worlds and Suns in Context" over the next decade. Currently this goal is very difficult with the current generation of 6.5-10m telescopes (on the ground or in space). The reason for this is simply that (from the ground-based telescope perspective) the best planet/star contrasts possible are in the $\sim 10^{-7}$-$10^{-8}$ range (such contrasts are possible even if the raw PSF contrasts are $10^{-5}$). While G-stars have habitable planet/star contrasts of $10^{-10}$, M-stars are much less (100-1000x) luminous, hence habitable planet/star ratios are closer to $10^{-7}$-$10^{-8}$. Hence, habitable rocky (0.5-2 $R_{earth}$) planets are, in theory, possible to detect around M-stars with ground-based telescopes[1-3]. However, the low luminosity of these M-stars means that the habitable zones (whose locations scale as $\sqrt{L_{star}}$) are 10-33x closer to their stars. At 5pc an earth-like planet around G2-star would have a habitable zone at ~200 mas (but contrasts of ~$10^{-10}$), whereas M-stars will have a habitable zone from ~6-20 mas (but much more reasonable contrasts of ~$10^{-7}$-$10^{-8}$). While these tiny (6-20 mas) separations are very difficult for today's 6.5-10m telescopes if we, however, look an ELT (like the D=25.4m Giant Magellan Telescope; GMT) then at $\lambda=0.8\mu m$ we find $3\lambda/D= 19.5$ mas, for the 30m TMT this is 16.5mas, and for the 40m ELT is 12 mas. So with the introduction of ELT sized apertures we have the possibility of directly detecting and characterizing reflected light (at ~$10^{-7}$ planet contrasts) from rocky planets in the habitable zones of nearby M-stars.

An excellent example of this type of target is the well-known radial velocity exoplanet Proxima Centauri b. This is closest exoplanet to the Sun (at just 1.3 pc) and is very likely a rocky 1.3 $M_{earth}$ planet in the habitable zone of its star. This planet reaches a maximum separation of 36 mas every 5.5 days, and due to its host star's M5.5 spectrum (just

0.0017 $L_{sun}$), the planet's contrast is a favorable ~$10^{-7}$. This is a great southern target for the GMT and ELT (TMT is too far north) in the search for biomarkers in the reflected light spectrum of potentially habitable planets. Prox. Cen. b is such a good target it might even be possible to detect with 6.5-10m visible wavelength ExAO like MagAO-X (phase II)[7].

For all the reasons listed above the *Astro2020* decadal survey listed characterization of potentially habitable planets around M stars as a key science goal for the ground-based ELTs. However, somewhat disappointingly, none of the first light instruments for the ELT, TMT, or GMT will be able to reach the contrasts and/or inner working angles (IWA) needed for this science. This is likely because it is assumed that the technology needed to reach IWAs of ~20mas and contrasts of ~$10^{-7}$ is beyond today's COTS technology for ELT sized telescopes. However, we have successfully completed an externally reviewed conceptual design (CoDR) for such an ELT scale instrument for the GMT telescope (called GMagAO-X) that should be capable of delivering this science at first light of the GMT[1-2] leveraging today's COTS technology.

Here we present the CoDR optical and mechanical design for the GMagAO-X instrument for Giant Magellan Telescope which is an ExAO system capable, in theory, of IWAs 20 mas and final planet contrasts of $10^{-7}$-$10^{-8}$ (with raw contrast of ~$10^{-5}$) at λ=0.8μm at 3λ/D, see Males et al. (2022)[4] for more on the science and performance of GMagAO-X. GMagAO-X should be able to detect and characterize many of the known, nearby, radial velocity (RV) planets in reflected light[1,4].

## 2. GMAGAO-X TOP LEVEL REQUIREMENTS

The Key top level requirement for GMagAO-X is to achieve a level of wavefront control on the 24.5m GMT similar to what is achieved by the current MagAO-X instrument on the 6.5m Magellan Telescope (Males et al. 2022[7]). This creates a flow down of high-level requirements on the opto-mechanics of GMagAO-X:

**T1.** To meet the fitting (14 cm sampling of primary) and servo error requirement, GMagAO-X must have 21,000 DM actuators capable of ≥2KHz correction speeds. See section 5 (E2E PERFORMANCE MODELING) of Males et al. (2022)[4] for a trade study proving the need for 21,000 actuators to hit a Strehl >80% at λ=0.8μm.

**T2.** To minimize wavefront/segment piston error we must have an interferometric beam combiner on a vibration isolated table.

**T3.** Must have stabilized GMT pupil on the Woofer and Tweeter DMs.

**T4.** It must have a coronagraph to suppress starlight

**T5.** It must have a post-coronagraphic focal plane wavefront sensor and Low Order WFS (LLOWFS) to sense non-common path (NCP) errors

**T6.** It must have a non-common path DM (NCP DM) to correct those NCP errors.

**T7.** It must do high-contrast science and wavefront sensing in the visible and the NIR.

### 2. 1 Optical mechanical compliance to the top level requirements

The high-level requirements above naturally flow down to the following optical and mechanical requirements and design decisions. The GMagAO-X Common-Path optical design allows for a number of critical OptoMech (OM) point design choices:

**OM1**. A beam reflected off M3 (vertex 1.000m above GIR) to travel 1.930 m to the f/8.157 focus (*compliant to GMT FP ICD*), *driven by T3*.

**OM2.** The beam travels to a gravity invariant (floating) optical table which is "stable" with only slow (<1 Hz) small (<50μm drifts). *Driven by T2*.

**OM3.** The rotating GMT pupil is stabilized to less than 1% of a tweeter DM actuator (1mrad, 0.06 deg) by a K-mirror, *closes out T3*.

**OM4.** The GMT M1 is first reimaged onto a commercial large stroke 93mm ALPAO 3228 actuator woofer DM (*compliant Woofer footprint 92x86.45mm*), *driven by T1*.

**OM5.** The pupil is then reimaged onto a "parallel DM" with seven commercial 3K BMC MEMS DMs (yielding a 21,000 actuator fast (>2KHz) Tweeter).

1. Each MEMS DM is 24.8mm dia. (62 actuators) but we only illuminate the inner 60-58 (*compliant tweeter beam 24x23.355mm/DM –outer segments slightly elliptical in shape*), *closes out T1*.
2. Fast >kHz segment piston control ±0.4μm OPD can be achieved with the unilluminated ring of outer actuators of each 3K Tweeter DM, *driven by T2*.

**OM6.** The seven BMC DMs use six PI 325 tip-tilt-piston actuated beam combiners (~100Hz ±42μm OPD dynamic range beam combiners) which are used to co-phase large, slow, phasing errors of the GMT from IR and/or Visible piston sensors. *Closes out T2*.

**OM7.** The upper optical table will have a 3k BMC DM for the removal of NCP errors. The upper table will have a coronagraph (PIAACMC baseline) and a reflective Lyot stop which feeds a LLOWFS. The LLOWFS will, in turn, control the NCP DM. *Closes out T4-T7*.

## 3. OPTICAL MECHANICAL DESIGN OF THE PARALLEL DM

### 3.1 Overview of the Parallel DM

The CoDR optomechanical design incorporates all OM1-OM7 design points. The most demanding of the design points was the top-level need T1 for 14cm sampling of the primary. This forces us to need a 21,000 actuator tweeter DM. This DM must also have many microns of segment tip-tilt and piston to correct for GMT segment vibrations and alignment errors. Such a DM is not available today, nor will one be available in the near (or far?) future.

We have invented a new type of tweeter DM, where the large problem of providing 21,000 actuators is split in a parallel problem of matching each of the seven 8.4m M1 segments with its own commercial 3,000 actuator (3k) DM. In this manner we have 7x 3k DMs which when used in parallel yield a 21,000 actuator tweeter that is available from COTS parts, we call this DM architecture a "parallel DM"[2].

| GMagAO-X Parallel DM (tweeter) Parameter | | value | |
|---|---|---|---|
| Number of commercial DMs | | seven BMC 3K 400μm pitch DMs | |
| Total # of Actuators | | 21,000 | |
| Stroke, speed | | 3.5μm, 3.5 kHz | |
| Unilluminated act. for fast (KHz) segment piston | Illuminated actuators | Outer ring unilluminated of each 3K DM | the center 24.0 mm dia., 60 act., are illuminated pupils, 73.013 mm metapupil |
| Approx. illuminated act. | | 19,700 | |
| Sampling of tweeter on GMT primary (M1) | | 14 cm | |
| Slow (100 Hz) Segment piston, and tip-tilt stroke | | ±42 μm, ±8 mrad (6x PI S-325 actuators in double pass) | |

Table 1: the GMagAO-X parallel design requirements and point design

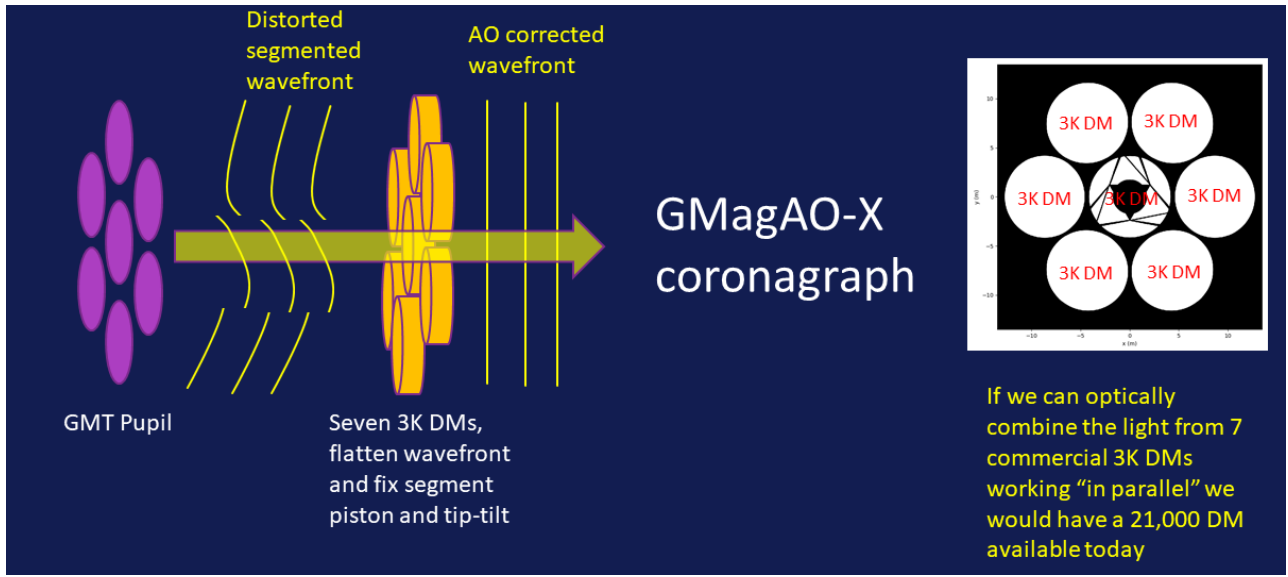

Fig 1: A cartoon motivating the parallel DM concept. Note that the parallel DM geometry is naturally well suited to fix segment tip-tilt and piston errors (no monolithic DM has 21,000 actuators and the ability to fix segment piston and tip-tilt).

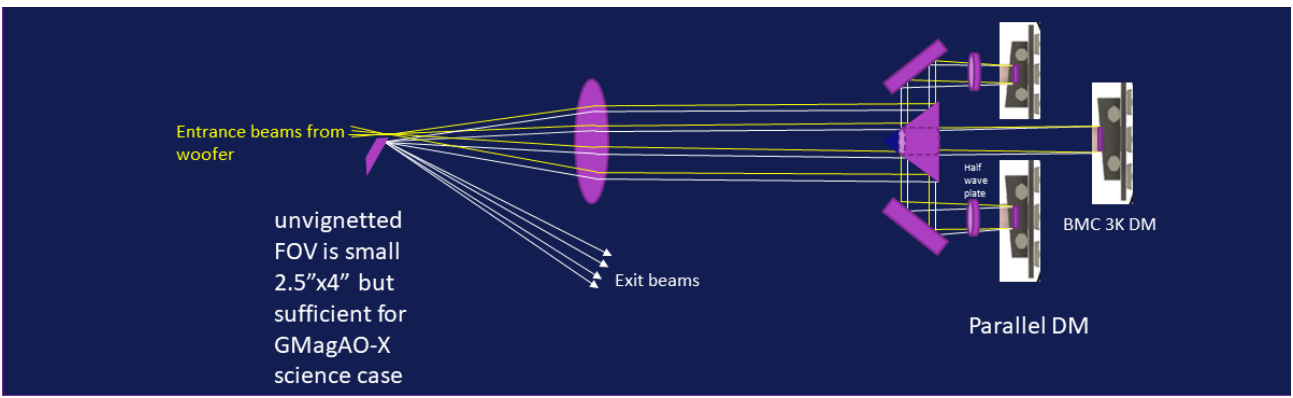

Fig 1: A cartoon of the parallel DM concept highlighting the pupil splitting by the hexpyramid[6] onto the commercial Boston Micromachines Corp (BMC) 3k DMs.

The opto-mechanical details of the parallel DM are simple in concept, but complex in detailed execution. Since this is a new concept, we will try to motivate our design with a few simple cartoons. As figure 1 illustrates we need an opto-mechanical way of slicing the seven 8.4m pupil segments of the GMT onto 7 individual "arms" that each reimage the 8.4m segment onto a 3k DM. As figure 2 shows it is possible to achieve this with the use of a central six sided reflective "hexpyramid". For the detailed opto-mechanical design of the hexpyramid please see Kautz et al. 2022[6]. As figure 3 illustrates it is also possible to actuate the fold flat with a commercial PI S-325 tip-tilt and piston actuator. This gives the DM a very large segment stroke. This parallel DM can tip-tilt each segment by ±8 mrad (optical) and ±42 μm of segment piston OPD at ~100 Hz (the allowed piston is $4/\sqrt{2}$ times the ±15μm of the 325 because it is at 45 deg AOI – verified in the HCAT lab). This is more than enough stroke and bandwidth to compensate for GMT segment vibrations and alignment errors.

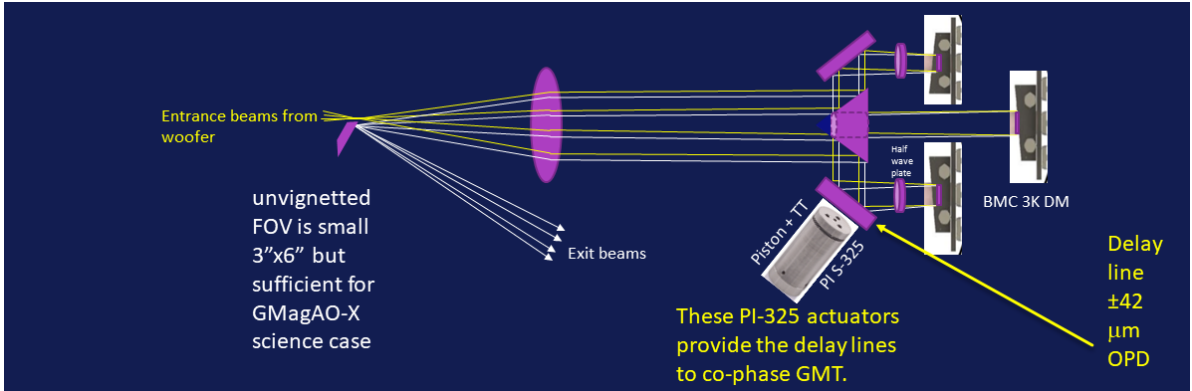

Fig 3: A cartoon of the parallel DM concept highlighting how the fold flats can be glued onto PI S-325 tip-tilt, piston actuators which can act as beam combiners (and segment tilt correctors) to keep the seven segments of the GMT co-phased. Due to this large piston stroke GMagAO-X can work with just the static (fast steering) secondaries of the GMT or it can use the adaptive secondaries when available.

### 3.2 The optical design for the parallel DM

As figs 2-3 make clear there are 5 reflections in the design (hexpyramid → fold flat → 3k DM → fold flat → hexpyramid) and we also need all seven beams to coherently combine. Unfortunately, the hexpyramid leads to seven different linear polarization angles of the returning beams, this limits the Strehl of the re-combined beam. Our solution is detailed in Hedglen et al. (2022)[8] where we show that "crossing" the fold mirror to reflect at 90 degrees to the hexpyramid allows the P→S polarization mode and S→P as well. In other words, if the hexpyramid and the fold flats are both at 45 deg AOI and they both have the same protective silver coating, then there will no significant linear polarization applied to any of the arms of the parallel DM. Detailed Zemax polarization models (see fig 4) of the "crossed-pair" parallel DM with our protective silver coatings show 100% Strehl (and zero induced linear polarization) at all wavelengths[8-9].

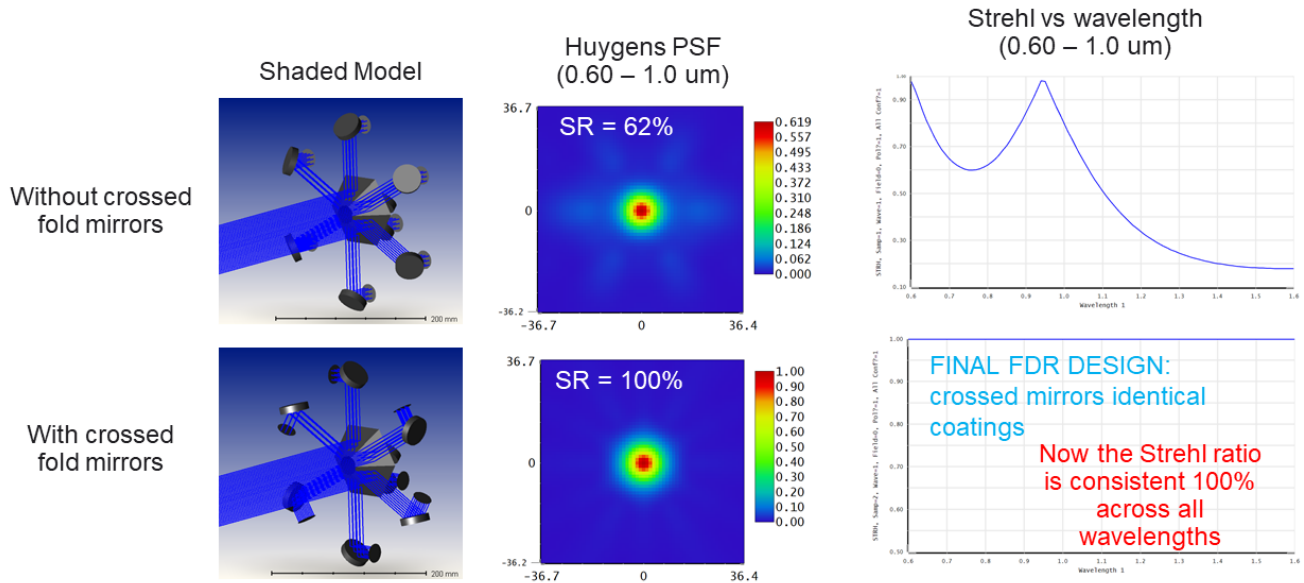

Fig 4: The final "crossed mirror" optical design that passed the HCAT testbed FDR review. This was fabricated on the HCAT testbed[6,8]. By crossing the fold flat at 90 deg w.r.t. the hexpyramid there is no loss of Strehl from polarization, Hedglen et al. (2022)[8].

### 3.3 The mechanical design for the parallel DM

Now that we have successfully found an optical solution for the parallel DM we can also have a mechanical design for holding the hexpyramid, six S-325s and the seven 3k DMs. Our solution is shown in figures 5 and 6.

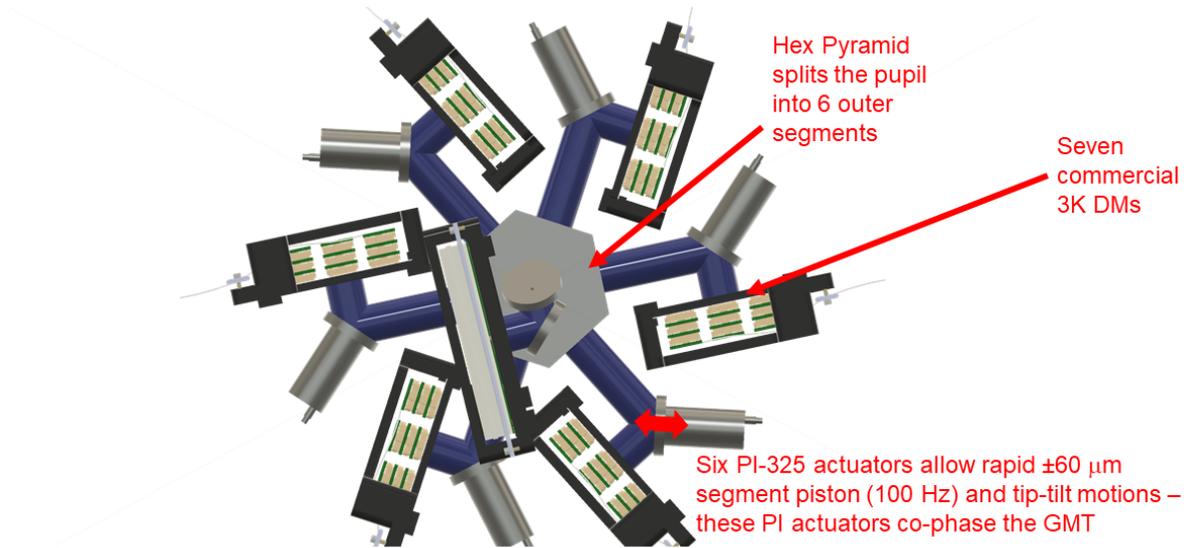

Fig 5: A Schematic of the GMagAO-X parallel DM, modified from Hedglen et al. (2022)[8].

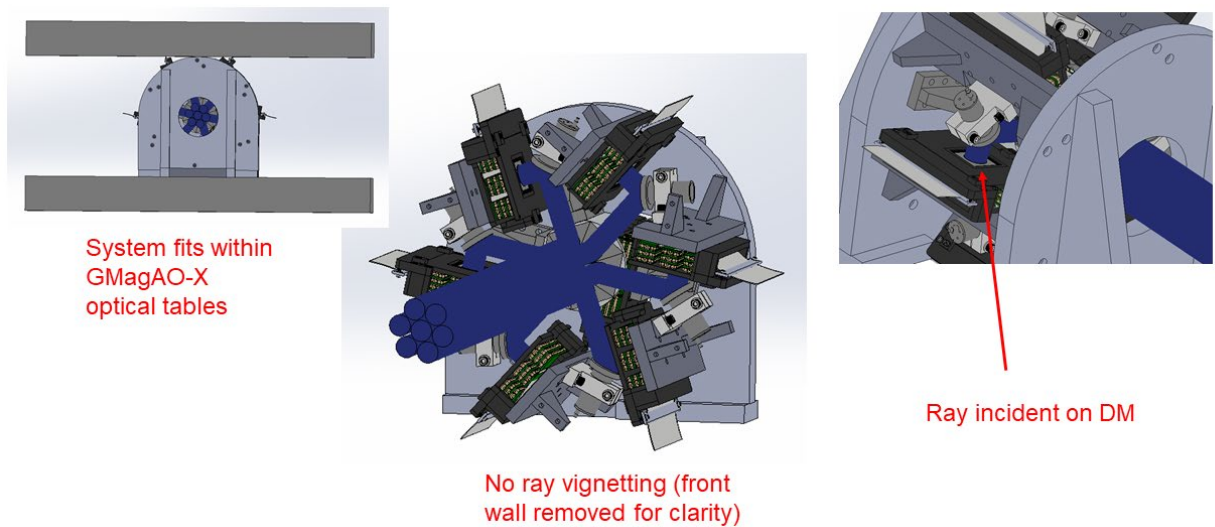

Fig 6: The mechanical design of the GMagAO-X parallel DM, modified from Kautz et al. (2022)[6].

As is clear from figure 6 and figure 4 the design is quite opto-mechanically complex. Due to the risk and novelty of this parallel DM design we were funded by the NSF risk reduction fund to the GMT (via AURA) to build a testbed version of the GMagAO-X parallel DM[12]. Since this version will be built at low cost the seven 3k BMC DMs were replaced by flat mirrors (which we call "mock DMs"). The rest of the testbed parallel DM is identical to that needed by GMagAO-X. Figure 7 shows the testbed version of the parallel DM that passed an external FDR review in Feb. 2022. The parallel DM will be fully tested in the very near future (Nov 2022)[6]. The HCAT testbed will feed a GMT pupil into the parallel DM and then passes that beam into the ExAO system MagAO-X when it is in the lab in Tucson[12]. See figure 8 from Hedglen et al. (2022)[8]. These tests will allow us to have confidence that we will be compliant for optomech requirements OM5 and OM6.

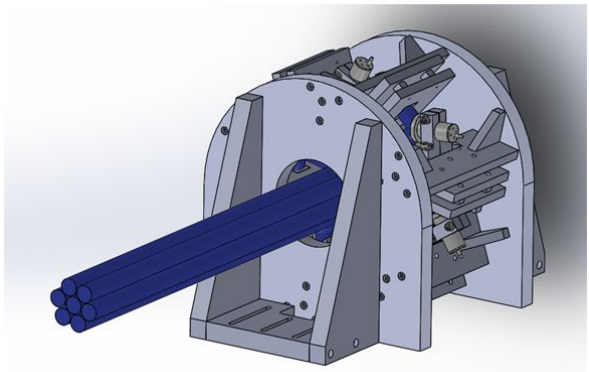
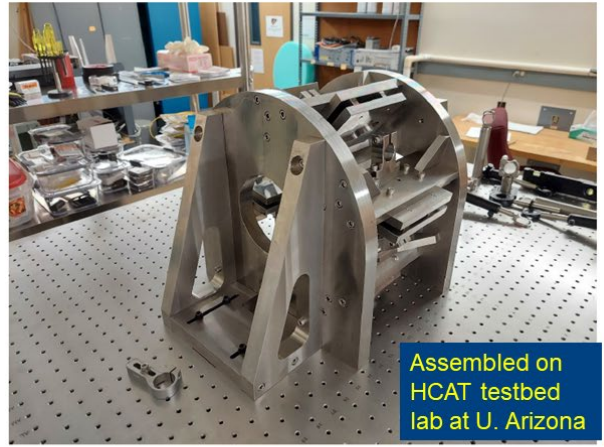

Fig 7: The as-built mechanical design of the GMagAO-X testbed parallel DM. See Kautz et al. (2022)[6] for more details and photos of the as-built invar parallel DM. *At this time we have successfully co-phased (white light fringes with HDFS) two beams of the parallel DM verifying our optical and mechanical design for a high Strehl recombined beam.* Soon all 7 beams will be recombined when the remaining 5x PI S-325s are delivered (delayed due to chip storages at PI).

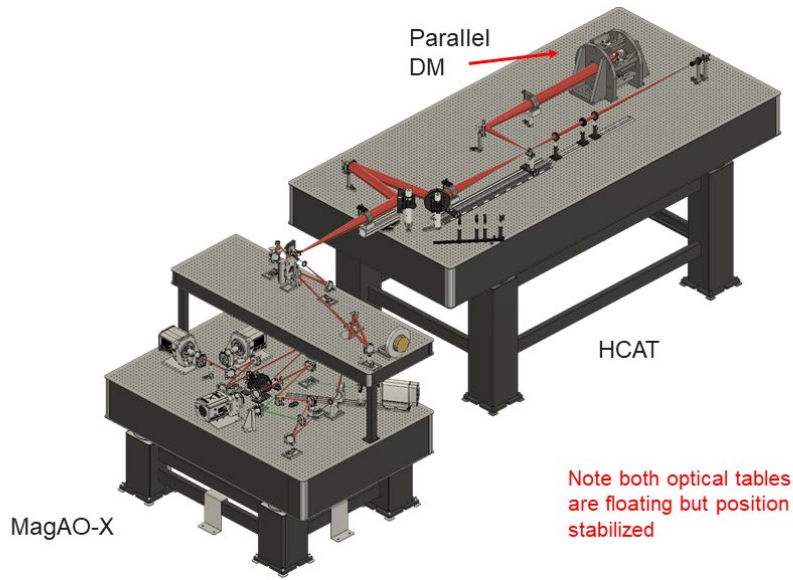

Fig 8: The HCAT testbed mechanics must provide a GMT-like pupil with 7 segments with 6 segments with rapid tip-tilt and piston. That pupil must be relayed to the MagAO-X 97 element woofer DM and 2040 actuator tweeter. It must also test the coherence of the parallel DM architecture needed for the GMagAO-X ExAO system. At this time all the optical/mechanical components shown above are in-house, except for one custom 4-inch triplet lens and the six PI E-727 controllers (due to chip shortages). All these components will be delivered and aligned by Nov 2022.

Based on successful tests of the parallel DM on the HCAT testbed, and future full funding for GMagAO-X, we will then move the parallel DM to the GMagAO-X instrument and add the seven 3k BMC DMs to fully populate it.

## 4. OPTOMECHANICAL DESIGN FOR GMAGAO-X

### 4.1 Mounting to the GMT

GMagAO-X must be on the top of the GIR platform at one of the 3 instrument folded ports. However, to minimize vibrations and flexure the GMagAO-X optical tables must be floating and gravity invariant (like those successfully used in our MagAO-X instrument[10]). However, unlike MagAO-X (which is on the gravity invariant Nasmyth platform of the 6.5m Magellan Clay telescope) the GIR platform floor tilts with the GMT's elevation axis. Our GMagAO-X operational model requires it to be rotated on the GIR to align with the elevation axis of the telescope (center of curvature of the C rings). GMagAO-X will also have a large derotator bearing that will counter-rotate the GMT elevation angle. In this manner, GMagAO-X's optical tables will always be upright w.r.t. gravity, hence the optics of GMagAO-X will be gravity invariant. Once gravity invariance is achieved the optical tables will be released from their clamps and floated into their operational position. We plan to use the same closed-loop servo floating optical system that has been successfully used with MagAO-X (TMC's PEPSII system), this architecture allows us to be compliant with OM1-OM3.

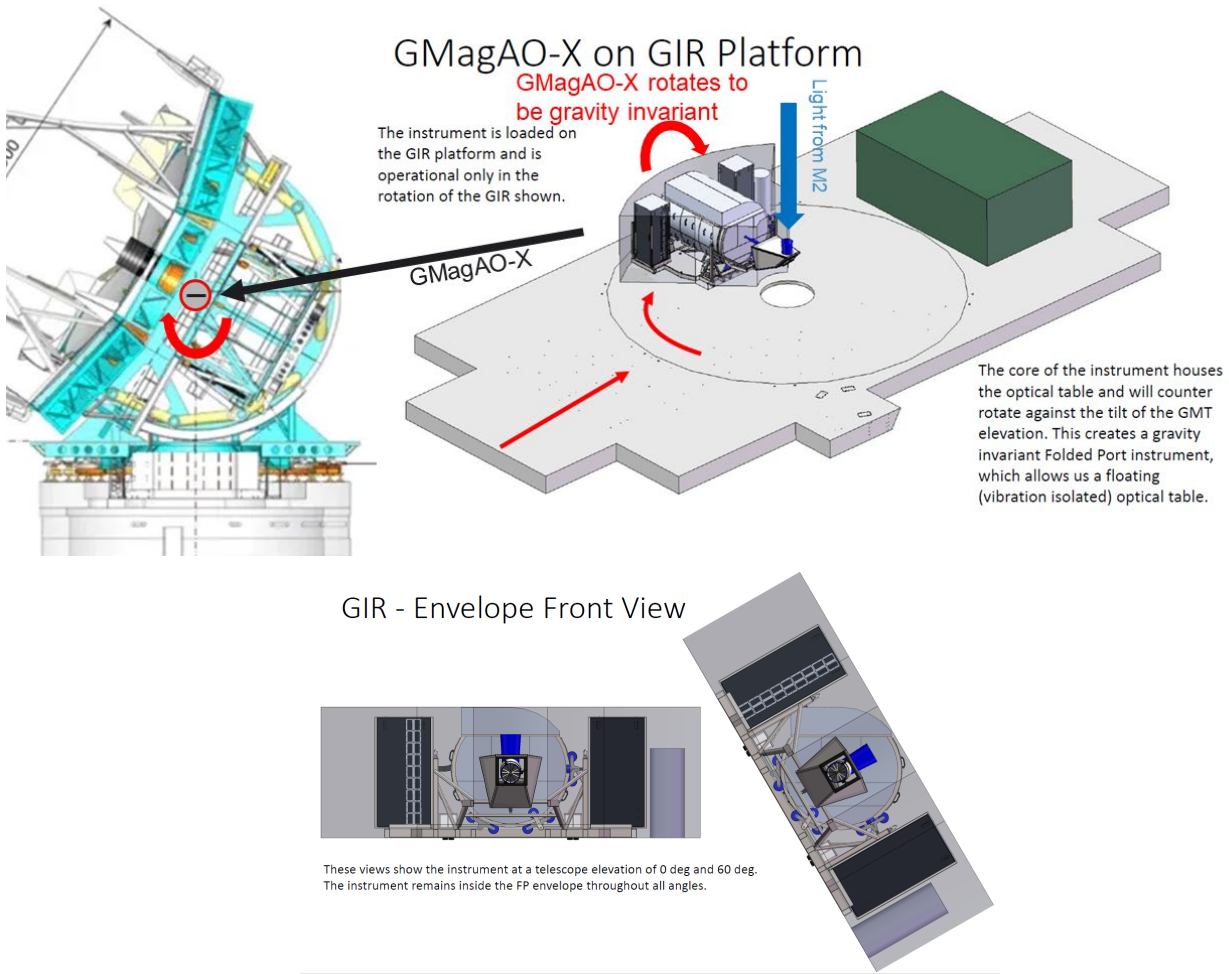

Fig 9: How GMagAO-X mounts to the GMT. Note how this provides a gravity invariant optical table. We plan to float these tables in the same manner as MagAO-X does. BOTTOM: Note how in the GIR front view image the "barrel" of the instrument is always pointed up (the flat roof is always on top), even if the GIR floor is angled at 60 degrees. This is how we keep GMagAO-X gravity invariant. GMagAO-X solid models are from the GMagAO-X CoDR produced by mechanical engineer Fernando Coronado.

## 4.2 The optical design of GMagAO-X

Figures 10-14 illustrate the CoDR design for the optics of GMagAO-X. The design goes from M3 through to the upper table's coronographic science cameras. For more on the scientific performance of GMagAO-X please see Males et al. (2022)[1]. The optical design was first done in python for maximum flexibility for optimizing the very compact folded design. Then the final design was built in ZMAX by optical engineer Oli Durney. The design is ~100% Strehl (assuming perfect optics) over our FOV (2.5x4").

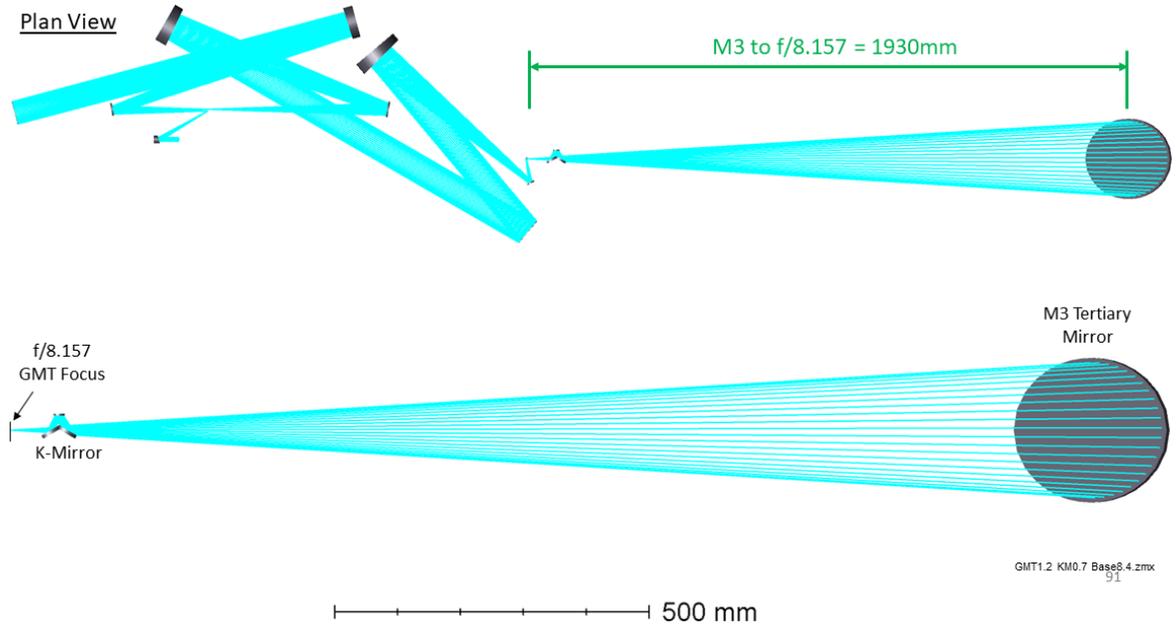

Fig 10: Common optical path from M3 to the GMT focal plane (bottom) and the full optical path of the lower table (at top).

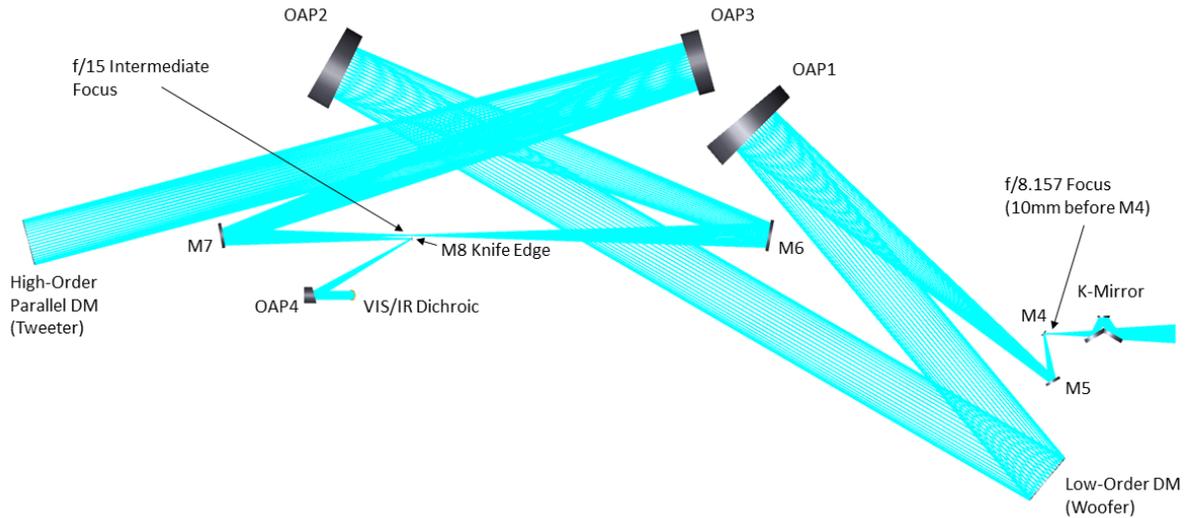

Fig 11: Zoom in on the common path from the K-mirror off the f/8.157 OAP1 to image the GMT pupil (M1) on the ALPAO 3228 woofer DM (satisfies OM4). Then the collimated beam is focused to an f/15 focal plane by OAP2, then the f/15 beam is re-collimated to the correct metapupil size and the GMT pupil is once again re-imaged on the tweeter DM. The tweeter DM is treated as just a flat mirror, even though it is in reality the full parallel DM shown in figures 4-6. The ADC (not shown) is just after the Woofer.

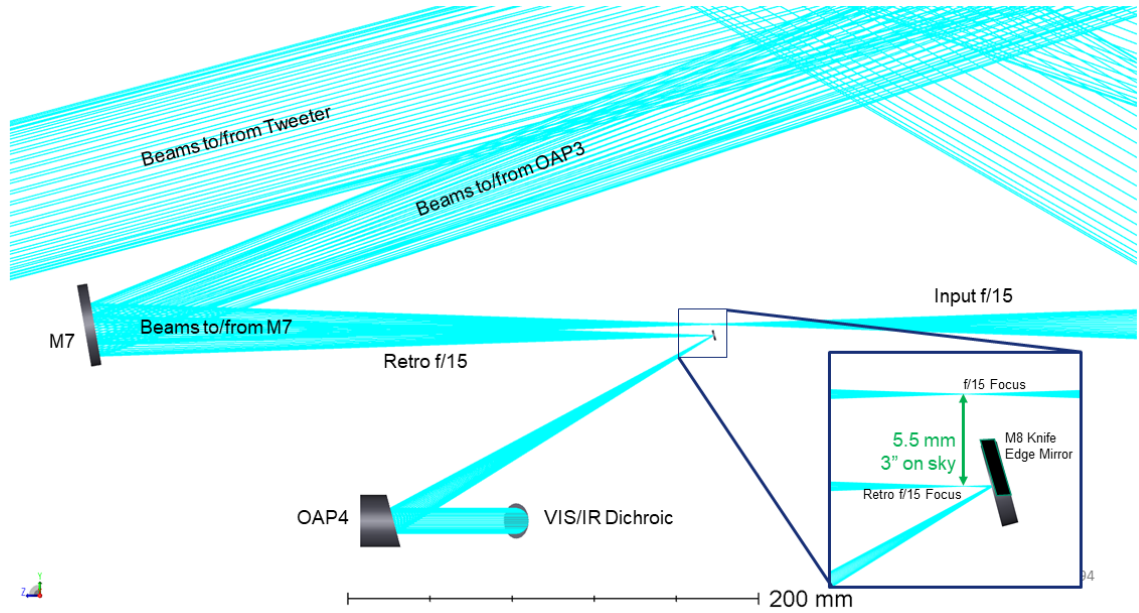

Fig 12: Close-up of the parallel DM's knife edge mirror to allow the system to be used in double pass. The FOV of this design is 2.5x4" on the sky (more then large enough for the GMagAO-X science cases). After OAP4 the light is split up to the IR HDFS (phasing sensor[11]) and the visible/NIR Zernike WFS. For more details about the HDFS segment phasing sensor see Haffert et al. (2022)[11]. A low order IR PyWFS drives the woofer, allowing >10% Strehl in the visible/NIR which allows the Zernike WFS to measure and correct the fine wavefront errors with the parallel DM Tweeter. For more about the GMagAO-X WFSs see Haffert et al. (2022)[5].

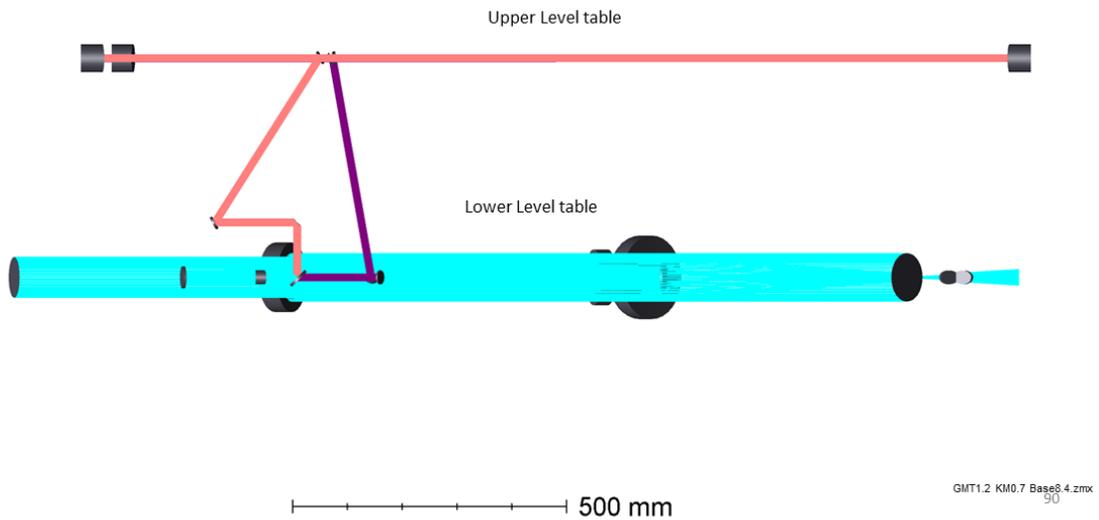

Fig 13: The side view of the optics, going from the lower table to the upper table where the coronagraphs and science cameras/spectrographs are located.

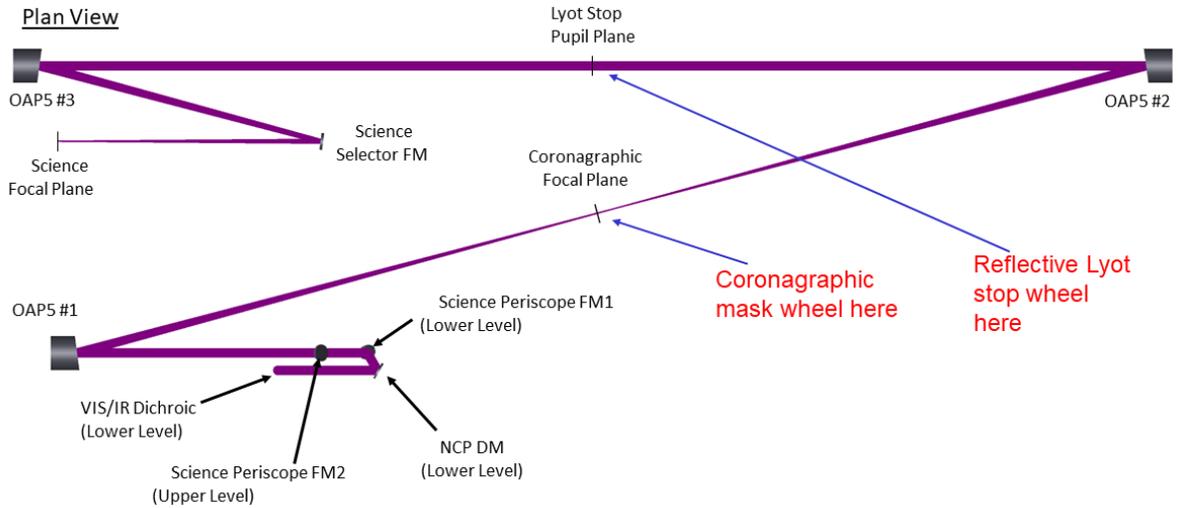

Fig 14: UPPER TABLE: The remaining Vis/IR light not used for the WFSs is now science light and is relayed up from the lower table. At a pupil there is an NCP DM that corrects any NCP errors measured in the science focal plane and/or the LLOWFS. The collimated beam is focused into an f/69 beam by OAP5#1 and then the coronagraphic focal plane id where a PIAACMC phase mask can be placed. The beam is collimated again at OAP5#2 and a pupil is focused on the Lyot stop. Since this stop is reflective, most of the starlight is then reflected into the LLOWFS camera off the reflective Lyot stop (not shown). Now most of the starlight is removed and OAP5#3 focuses the f/69 beam onto the science camera, or the "science selector" flat can steer the beam to an IFU fiber feed to another instrument (like, for example, the G-CLEF spectrograph). This upper table design satisfies optomech requirement OM7.

### 4.3 The Mechanical Design

Now we can see the ZEMAX lower table optical design (highlighted in figures 10-12) incorporated into the solidworks mechanical design in figs. 15-20. GMagAO-X is (with ~10% margin applied) 6,268kg (below the required 6,400 kg) and fits into the required space envelope of a GMT FP instrument. We are also compliant with all the other GMT FP ICDs.

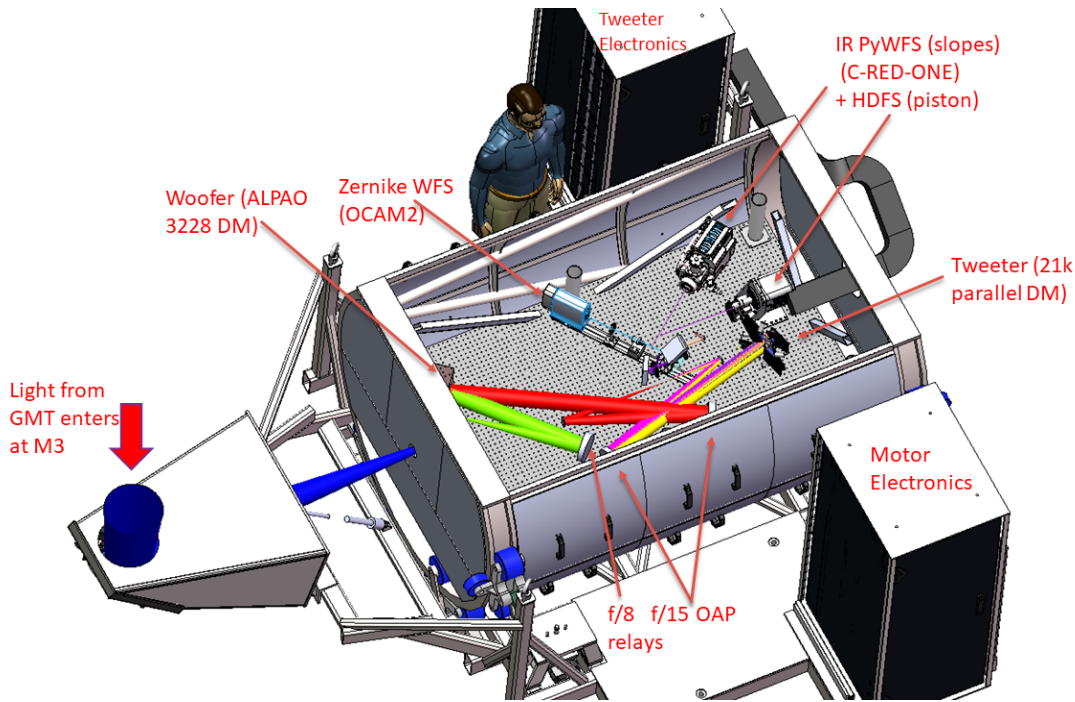

Fig 15: The optomechanical design showing the deployed M3 and the lower table optics (the upper table is removed for clarity).

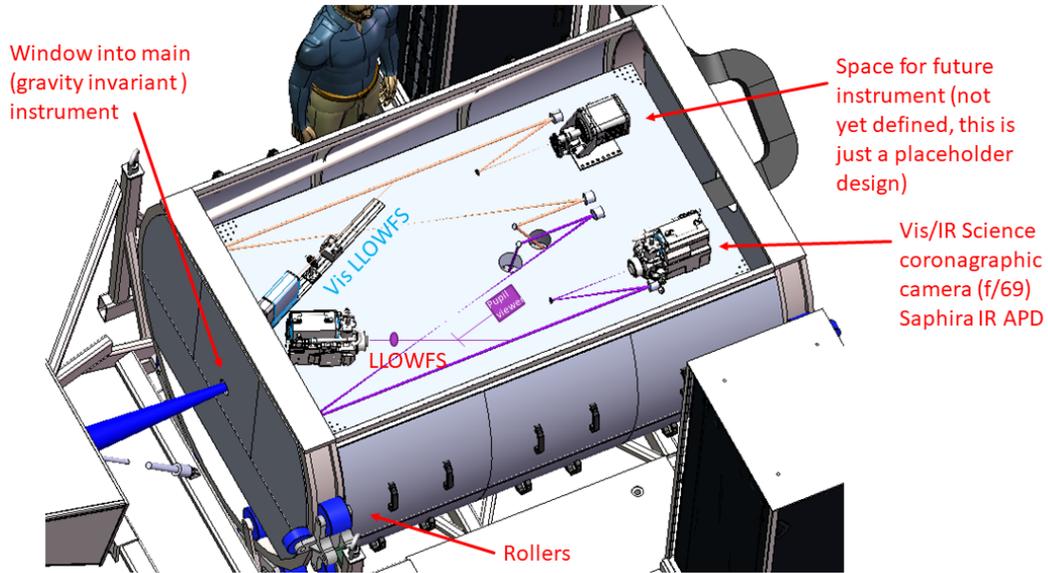

Fig 16: The optomechanical design showing the upper optical table, coronagraphs, LLOWFS (to sense any NCP errors), and science cameras. The roof of the instrument has been removed for clarity.

To get a better understanding of how the double decker optical tables float inside the de-rotating "barrel" of GMagAO-X we have a cut-away image of the instrument cut exactly through the middle of the solid model in figure 17.

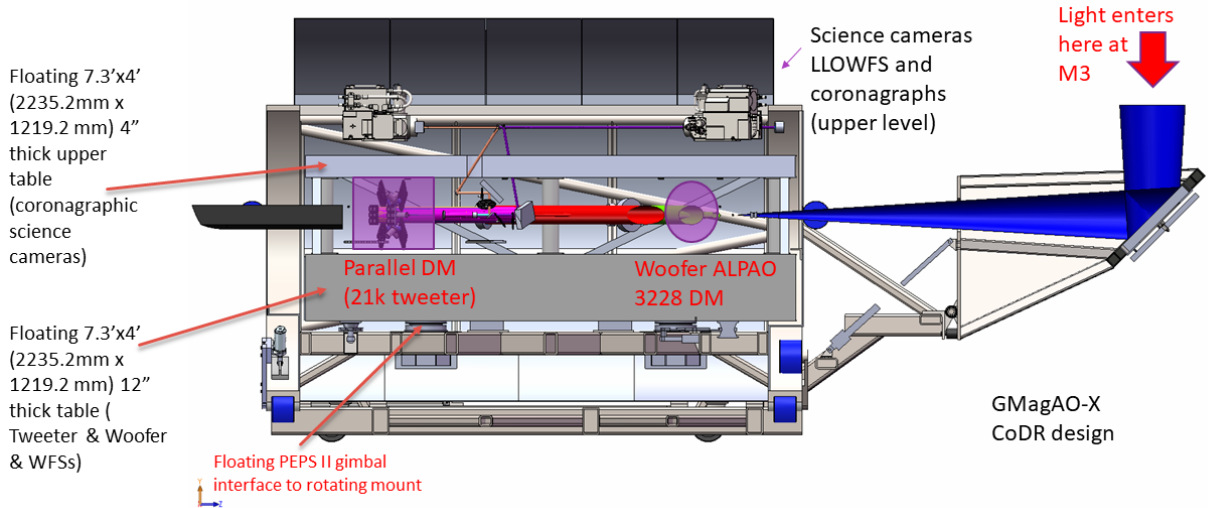

Fig 17: The optomechanical design cross section of GMagAO-X.

## 5. NEXT STEPS

We will finish proving/testing the parallel DM concept with our full scale parallel DM on the HCAT testbed in the lab by Nov 2022[6]. We will bring the whole instrument up to the PDR level in 2023[4]. The optomechanics of the science instruments will be further developed by PDR. The PIAACMC coronagraph will be better integrated into the optomechanical models at PDR. The GMagAO-X WFS will have a full optical design by PDR as well[5]. A stiffer design for the M3 mount will also be investigated with a new FEA analysis of alternate support systems from the GIR floor.

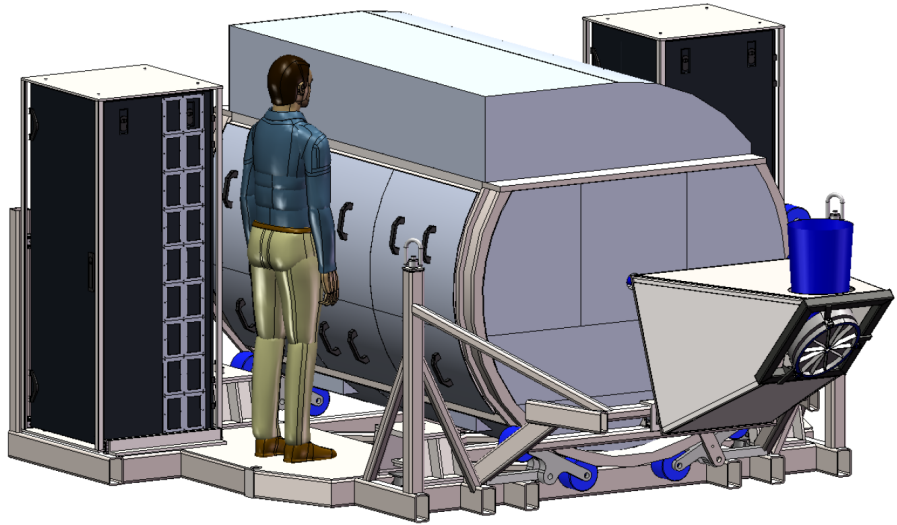

Fig 18: An iso-view of the GMagAO-X instrument with its dust covers on. Six foot human for scale.

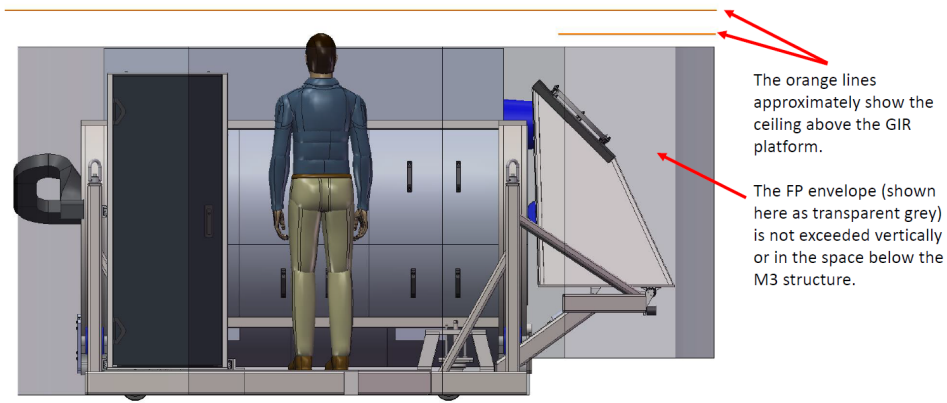

Fig 19: A side view of GMagAO-X when not in use (M3 is retracted completely out of the GMT beam). The gray shaded/transparent envelope is the GMT Folded Port (FP) instrument envelope. GMagAO-X fits completely inside the required FP envelope.

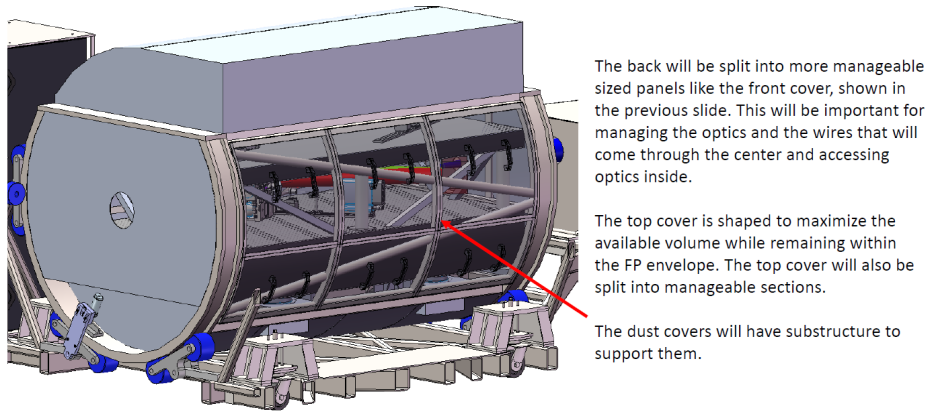

Fig 20: A detail image showing how the dust panels for the instrument can be removed to allow access to the optics, etc.

## 6. CONCLUSIONS

Here we presented the CoDR optomechanical design (which successfully passed an external CoDR review in September 2021) for the ExAO instrument GMagAO-X. GMagAO-X is designed for the high level goal to reach ~$10^{-7}$ contrasts at ~20 mas. Such an instrument should enable reflected light characterization of many known RV exoplanets, some in the habitable zones of their stars. In particular, the potentially habitable planet Prox. Cen. b would be an excellent target for biomarker searches with GMagAO-X. However, to reach these goals there is a requirement to have a 21,000 Tweeter DM (significantly larger than any available today). We show that a parallel DM with seven commercial BMC 3k DMs will provide all the stroke and speed required (as well as the segment tip-tilt and piston phasing). The rest of the GMagAO-X CoDR design also utilizes essentially COTS parts. Hence, there are no items in the design that require external technical development. However, there is still significant room for improvement in the design as we move to PDR. In particular, improvement on the mechanical design of the M3 mount and maturing the WFS design and optomechanics will be done by PDR. Nevertheless, our positively reviewed CoDR design shows that GMagAO-X is a very viable ELT-scale ExAO system that could be operational at, or near, first light of the GMT.

## ACKNOWLEDGMENTS

The HCAT testbed program is supported by an NSF/AURA/GMTO risk-reduction program contract to the University of Arizona (GMT-CON-04535, Task Order No. D3 High Contrast Testbed (HCAT), PI Laird Close). The authors acknowledge support from the NSF Cooperative Support award 2013059 under the AURA sub-award NE0651C. Support for this work for SYH was also provided by NASA through the NASA Hubble Fellowship grant #HST-HF2-51436.001-A awarded by the Space Telescope Science Institute, which is operated by the Association of Universities for Research in Astronomy Inc. (AURA), under NASA contract NAS5-26555. Maggie Kautz received an NSF Graduate Research Fellowship in 2019. Alex Hedglen received a University of Arizona Graduate and Professional Student Council Research and Project Grant in February 2020. Alex Hedglen and Laird Close were also partially supported by NASA eXoplanet Research Program (XRP) grants 566 80NSSC18K0441 and 80NSSC21K0397 and the Arizona TRIF/University of Arizona "student link" program. We are very grateful for support from the NSF MRI Award #1625441 (for MagAO-X). This material is based upon work supported in part by the National Science Foundation as a subaward through Cooperative Agreement AST-1546092 and Cooperative Support Agreement AST-2013059 managed by AURA. We are very grateful for the University of Arizona Space Institute for funding the past CoDR and future PDR efforts for GMagAO-X (PI Jared Males).